\documentclass[letterpaper]{article} %
\usepackage{aaai24}  %
\usepackage{times}  %
\usepackage{helvet}  %
\usepackage{courier}  %
\usepackage[hyphens]{url}  %
\usepackage{graphicx} %
\usepackage{natbib}  %
\usepackage{caption} %
\usepackage{algorithm}
\usepackage{algorithmic}

\usepackage{newfloat}
\usepackage{listings}
\DeclareCaptionStyle{ruled}{labelfont=normalfont,labelsep=colon,strut=off} %
\floatstyle{ruled}
\newfloat{listing}{tb}{lst}{}
\floatname{listing}{Listing}
\usepackage[utf8]{inputenc}
\usepackage{amssymb}
\usepackage{tabto}
\usepackage{amsthm}
\usepackage[english]{babel}
\usepackage{mdframed}
\usepackage{enumitem} %
\usepackage{comment}
\usepackage{tabularx}
\usepackage{tikz} %
\usetikzlibrary {automata,positioning} 
\usepackage{stmaryrd} %
\usepackage{bbold} %
\usepackage{bbding} %
\usepackage{multicol} %
\usepackage{subcaption} %
\usepackage{float} %
\usepackage{mathtools} %
\usepackage[color=lime]{todonotes}
\usepackage{subcaption}
\usepackage{color}

\usepackage{environ} %
\usepackage{apxproof} %

\usepackage{pifont} %
\newcommand{\cmark}{\ding{51}}
\newcommand{\xmark}{\ding{55}}

\newtheoremstyle{rq}%
	{1pt}%
	{1pt}%
	{\upshape}%
	{}%
	{\bfseries}%
	{ :}%
	{.5em}%
	{}%
	
\newtheoremstyle{axiom}%
	{1pt}%
	{1pt}%
	{\upshape}%
	{}%
	{\bfseries}%
	{ : \\}%
	{\newline}%
	{}%

\newtheoremstyle{mydef}%
	{10pt}%
	{10pt}%
	{\upshape}%
	{}%
	{\bfseries}%
	{.}%
	{.5em}%
	{}%

\newtheoremstyle{mytheorem}%
	{10pt}%
	{10pt}%
	{}%
	{}%
	{\bfseries}%
	{.}%
	{.5em}%
	{}%

\theoremstyle{mytheorem}
\newtheoremrep{thm}{Theorem}
\newtheoremrep{prop}[thm]{Proposition}
\newtheoremrep{lem}[thm]{Lemma}
\newtheoremrep{cor}[thm]{Corollary}
\newtheoremrep{ex}[thm]{Example}
\newtheoremrep{conj}[thm]{\textcolor{teal}{Conjecture}}

\theoremstyle{rq}

\theoremstyle{mydef}
\newtheorem{defin}[thm]{Definition}

\newcommand{\ltuple}{\left(}
\newcommand{\rtuple}{\right)}

\newcommand{\coalition}{C}
\newcommand{\sbar}{\bar{s}}
\newcommand{\Sbad}{\lightning} %

\newcommand{\ts}{\mathit{TS}}
\newcommand{\tsStates}{S} %
\newcommand{\tsTrans}{\to} %
\newcommand{\tsInit}{s_0} %
\newcommand{\tsTuple}{\ltuple \tsStates, \tsTrans, \tsInit, \Sbad \rtuple}

\newcommand{\tsRun}{\rho} %
\newcommand{\ce}{\rho} %

\newcommand{\coopGamePlayers}{X} %
\newcommand{\coopGame}{v} %
\newcommand{\coopGames}[1][\coopGamePlayers]{G^{#1}} %
\newcommand{\criticalPair}[2]{\ltuple #1, #2 \rtuple}
\newcommand{\ws}{\mathtt{WS}_\optimistic} %
\newcommand{\rs}[1]{\mathtt{RS}_\optimistic(#1)} %

\newcommand{\semivalueName}{general power index}
\newcommand{\semivaluesName}{general power indices}

\newcommand{\SemivaluesName}{General power indices}

\newcommand{\weightVec}{p}

\newcommand{\weightVecTuple}[1]{\ltuple \weightVec_0, \ldots, \weightVec_{#1} \rtuple}
\newcommand{\semivalue}{\mathcal{R}}
\newcommand{\weights}{\mathtt{Weights}}

\newcommand{\optimistic}{\mathtt{opt}}
\newcommand{\pessimistic}{\mathtt{pes}}

\newcommand{\vopt}{\coopGame_{\optimistic}} %
\newcommand{\vpes}{\coopGame_{\pessimistic}} %
\newcommand{\voptfull}{\coopGame_{\optimistic}^{\ts, \rho}} %
\newcommand{\vpesfull}{\coopGame_{\pessimistic}^{\ts, \rho}} %
\newcommand{\vfull}{\coopGame^{\ts, \rho}} %

\newcommand{\shap}{\mathcal{S}} %
\newcommand{\ban}{\mathcal{B}} %

\newcommand{\pSafe}{\textit{Safe}}
\newcommand{\pReach}{\textit{Reach}}
\newcommand{\sSafe}{S_{\pSafe}}
\newcommand{\sReach}{S_{\pReach}}
\newcommand{\sAll}{S}
\newcommand{\sInit}{s_0}
\newcommand{\gameTrans}{\to}

\newcommand{\Srho}{\rho} %

\newcommand{\tsGameSymb}{\mathcal{G}}
\newcommand{\tsGame}[1]{\mathcal{G}^{TS}_{\rho}\left(#1\right)} %
\newcommand{\coalitionOpt}[1][C]{#1 \cup (\sAll \setminus \Srho)}

\newcommand{\switch}{\mathtt{Swi}}

\title{Backward Responsibility in Transition Systems Using General Power Indices}
\author{
	Christel Baier\textsuperscript{\rm 1,2},
	Roxane van den Bossche\textsuperscript{\rm 3},
	Sascha Klüppelholz\textsuperscript{\rm 1}, \\
	Johannes Lehmann\textsuperscript{\rm 1,2},
	Jakob Piribauer\textsuperscript{\rm 1} \footnote{Authors are listed in alphabetical order}
}
\affiliations{
	\textsuperscript{\rm 1}TU Dresden, Germany\\
	\textsuperscript{\rm 2}Centre for Tactile Internet with Human-in-the-Loop (CeTI) \\
	\textsuperscript{\rm 3}Université Paris-Saclay, ENS Paris-Saclay, France\\
	\{christel.baier, sascha.klueppelholz, johannes\_alexander.lehmann, jakob.piribauer\}@tu-dresden.de, roxane.van\_den\_bossche@ens-paris-saclay.fr
}

\begin{document}

\maketitle

\begin{abstract}
	To improve reliability and the understanding of AI systems, there is increasing interest in the use of formal methods, e.g. model checking. Model checking tools produce a counterexample when a model does not satisfy a property. Understanding these counterexamples is critical for efficient debugging, as it allows the developer to focus  on the parts of the program that caused the issue.
	
	To this end, we present a new technique that ascribes a responsibility value to each state in a transition system that does not satisfy a given safety property. The value is higher if the non-deterministic choices in a state have more power to change the outcome, given the behaviour observed in the counterexample. For this, we %
	employ a concept from cooperative game theory -- namely \semivaluesName{}, such as the Shapley value -- to compute the responsibility of the states.
	
	We present an optimistic and pessimistic version of responsibility that differ in how they treat the states that do not lie on the counterexample. We give a characterisation of optimistic responsibility that leads to an efficient algorithm for it and show computational hardness of the pessimistic version. We also present a tool to compute responsibility and show how a stochastic algorithm can be used to approximate responsibility in larger models. These methods can be deployed in the design phase, at runtime and at inspection time to gain insights on causal relations within the behavior of AI systems.
\end{abstract}

\section{Introduction}

Due to the ever-growing demand for reliable, trustworthy AI systems, research on ways to incorporate formal methods for the verification of such systems is becoming more and more important (see e.g. \citet{seshia2022toward}). AI systems are in general instances of parallel systems that operate in complex environments. As such, techniques such as model checking are suitable for this setting. The aim of model checking is to analyse a given system to prove automatically that this system satisfies some property. For example, we might require certain states to never be reached, because they correspond to an undesired event. When a model does not satisfy a safety property, the model checker returns a \emph{counterexample}, which is an execution of the model that violates the property (see \citet{Baier2008} for more details). 
While this demonstrates that an error exists, it is not obvious which part of the model is \emph{responsible} for the error. Some states in the counterexample may be unable to change the outcome, whereas others can ensure that the safety property is satisfied. 

We distinguish between \emph{forward} and \emph{backward} responsibility \cite{ForwardBackwardRelation}. In the forward case, responsibility depends only on the model, whereas in the backward case, it also takes a specific counterexample into account. Existing causality-based approaches to backward responsibility include the use of distance metrics \cite{ErrorsDistanceMetrics}, mutation-based techniques \cite{beer2012explaining}, event-order logic \cite{CausalityCheckingForComplexSystemModels} and hyperproperties \cite{TemporalCausality}.

In this paper, we introduce an intuitive, game-based approach to compute \emph{backward} responsibility. Our approach is inspired by \citet{ForwardResponsibility} where the Shapley value \cite{ShapleyValueOriginal} is used to determine a numeric value of \emph{forward} responsibility. For a more recent discussion of the Shapley value, we refer the reader to \citet{ShapleyDetails}. The Shapley value is also used for responsibility attribution by \citet{ResponsibilityGameTheory}, who present a technique that operates on game trees.

Incorporating a counterexample allows us to analyse a specific fault in the system. Moreover, we extend our definitions to \semivaluesName{}, also known as semivalues \cite{Semivalues}, of which the Shapley value is an instance.

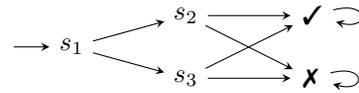
\begin{figure}[H]
    \centering
    \begin{tikzpicture}[>=stealth,node distance=1.1cm,auto, state/.style={circle,inner sep=2pt}]
        \node[state,initial,initial text=]  (one) {$s_1$};
        \node[state] (two) [above right = 0cm and 1.1cm of one] {$s_2$};
        \node[state] (three) [below right = 0cm and 1.1cm of one] {$s_3$};
        \node[state] (destination) [right = of two] {\cmark};
        \node[state] (accident) [right = of three] {\xmark};
    
        \path[->]
            (one) edge node {} (two)
            (one) edge node {} (three)
            (two) edge node {} (destination)
            (two) edge node {} (accident)
            (three) edge node {} (destination)
            (three) edge node {} (accident)
            (destination) edge [loop right] node {} (destination)
            (accident) edge [loop right] node {} (accident);
        
    \end{tikzpicture}
	\caption{Railway network with three switches $s_1, s_2\text{ and }s_3$. If a train is routed to \xmark{}, an accident occurs.}
    \label{fig:train_example}
\end{figure}

To illustrate our scenario, consider the railway network depicted in Figure~\ref{fig:train_example} with three switches $s_1$, $s_2$ and $s_3$. The goal is to route the train to the destination \cmark. However, it can also be routed to the unfinished line \xmark, which causes an accident. We investigate how the switches share the responsibility for the outcome.

In the forward responsibility, $s_2$ and $s_3$ would have the same responsibility due to the symmetry of the system. Now consider a scenario where an accident has occurred after the train took the path $s_1 s_2 \text{\xmark}$. It is now desirable to ascribe more responsibility to $s_2$ than to $s_3$.

To compute the exact responsibility, we determine which coalitions of states can ensure that \xmark{} is not reached. For example, given that the train travelled from $s_1$ to $s_2$, $\{s_2\}$ suffices: if it routes the train to \cmark{} instead of \xmark, the accident is averted. For $\{s_1\}$, the answer is less clear. It can route the train to $s_3$ instead, but we have not observed the behaviour of $s_3$ in the system. We therefore distinguish between two types of responsibility: For \emph{optimistic responsibility}, we assume that these states help us in trying to satisfy the safety property. For \emph{pessimistic responsibility}, on the other hand, we assume that they attempt to violate the property. For optimistic responsibility, $\{s_1\}$ can avoid \xmark, whereas for pessimistic responsibility, it cannot (but $\{s_1, s_3\}$ can). We analyse both types of backward responsibility in this article.

\textbf{Main Contributions: } The main contributions of this paper are: We give a definition of backward responsibility based on \citeauthor{ForwardResponsibility} and define the novel notion of optimistic responsibility in Section~\ref{sec:responsibility}. We analyse the complexity of pessimistic and optimistic responsibility in Section~\ref{sec:complexity}. We provide an implementation, show how a stochastic algorithm can be used to analyse large models and evaluate our implementation with several experiments in Section~\ref{sec:implementation}.

\textbf{Other Related Work: } There are several other approaches to assess backward responsibility. In addition to a counterexample, many of these require one or more passing runs, which are then compared. An example of this is \emph{Delta debugging} \cite{DeltaDebugging}.  \citet{LocalisingErrorsInCounterexampleTraces} take an intraprocedural approach and also use the information to identify multiple errors that may be present in a single model. \citet{ErrorsDistanceMetrics} use distance metrics to identify the differences between passing runs and a failing run and present them to the user as an explanation. A similar approach using nearest-neighbor queries is presented by \citet{FaultsNearestNeighborQueries}.

On the other hand, the technique of \citet{WhodunitCausalCounterexampleAnalysis} does not require additional runs and instead analyses the counterexample using weakest-precondition reasoning.

Similar to our technique and to \citet{ForwardResponsibility}, \citet{ResponsibilityGameTheory} use the Shapley value and game theory to analyse responsibility. They operate on games in tree form and also distinguish between forward and backward responsibility.

A related concept is \emph{blameworthiness} as defined by \citet{Blameworthiness} is a notion aiming to define \emph{moral} responsibility of agents. Our notion of responsibility does not include any moral considerations. 
Nevertheless, notions of blameworthiness can be defined using the technical vehicle of power indices as used for our notion of responsibility, such as the Shapley value used in
\citet{BlameworthinessMultiAgent}.

\section{Preliminaries}
\label{sec:preliminaries}

\emph{Transition systems.}
A \textit{transition system} is a tuple $\ts = \tsTuple$ where $\tsStates$ is a finite set of  \textit{states}, $\tsTrans$ is the \emph{transition relation} on $\tsStates$, $\tsInit \in \tsStates$ is the \textit{initial state} and $\Sbad \subseteq \tsStates$ is the set of \emph{bad states}. A \textit{run} on $\ts$ is an infinite sequence of states $\tsRun = \tsRun_0 \tsRun_1 \ldots \in \tsStates^\omega$, where $\rho_0 = \tsInit$ and $\forall i \in \mathbb N$ $\tsRun_i \tsTrans \tsRun_{i+1}$.

We call $\tsRun = \tsRun_0 \ldots \tsRun_k \in \tsStates^\ast$ a counterexample if it is the prefix of a run, $\tsRun_k \in \Sbad$, $\tsRun_i \not\in \Sbad$ for all $i \in \{0, \ldots, k-1\}$ and $\tsRun_i \neq \tsRun_j$ for all $i \neq j$, i.e. they are loop-free.

\emph{Cooperative games.}
Let $\coopGamePlayers$ be a finite set of \emph{players}. A \textit{cooperative game} on $\coopGamePlayers$ is a function $\coopGame \colon 2^{\coopGamePlayers} \to \mathbb{R}$ that associates a value to each subset of $\coopGamePlayers$. We call $C \subseteq X$ a \emph{coalition} and $\coopGame(\coalition)$ the \emph{gain} of the coalition $\coalition$. The set of games on $\coopGamePlayers$ is denoted by $\coopGames$.

We call $\coopGame$ \emph{simple} if $\coopGame(\coalition) \in \{0, 1\}$ for all $\coalition \subseteq 2^{\coopGamePlayers}$. A coalition $C \subseteq 2^{\coopGamePlayers}$ is \emph{winning} if $\coopGame(\coalition) = 1$. $\criticalPair{\coalition}{s}$ forms a \textit{critical pair} if $\coalition \cup \{s\}$ is winning and $\coalition$ is not.

\emph{\SemivaluesName.}
We call $\weightVec = \weightVecTuple{n-1}$ a \emph{weight vector} if
\begin{center}
    $\displaystyle \sum_{k=0}^{n-1} \binom{n-1}{k}\weightVec_k =1$.
\end{center}

Let $\coopGamePlayers$ be a finite set of players with $n \coloneqq \left| \coopGamePlayers \right|$. Then $\semivalue\colon \coopGames \to X \to \mathbb R$ is a \emph{\semivalueName{}} if there exists a weight vector $\weightVec = \weightVecTuple{n-1}$  such that, for any game $\coopGame \in \coopGames$ and player $i \in \coopGamePlayers$, we have

\begin{center}
    $\semivalue(\coopGame, i) = \displaystyle \sum_{\coalition \subset \coopGamePlayers \setminus \{i\}} \weightVec_{|\coalition|}[\coopGame(\coalition\cup \{i\})-\coopGame(\coalition)]$
\end{center}
where we write $\semivalue(\coopGame, i)$ instead of $(\semivalue(\coopGame))(i)$.

We note that for every \semivalueName{} $\semivalue$, only one such weight vector $\weightVec = \weightVecTuple{n-1}$ exists and define $\weights_i(\semivalue) = p_i$.

This definition corresponds to the characterisation of semivalues for games on finite sets of players given by \citet{Semivalues}. The linear games they use are isomorphic to functions from $\coopGamePlayers$ to $\mathbb{R}$, so our definition is equivalent to their definition and characterisation.

\textit{Shapley value, Banzhaf value.} Let $\coopGamePlayers$ be a finite set of players and let $\coopGame$ be a game on $\tsStates$. The \textit{Shapley value} is the \semivalueName{} $\shap$ with $\weights_i(\shap) = \frac{(n-i-1)!i!}{n!}$. The \textit{Banzhaf value} is the \semivalueName{} $\ban$ with $\weights_i(\ban) = \frac{1}{2^{n-1}}$.

We use the following well-known characterisation of the Shapley value in several proofs.%
\begin{lem}
\label{eqDefShapley}
    Let $X$ be a finite set of players. We have
    $\shap(v,x) = \sum_{\pi \in \Pi_X}\left(\coopGame(\pi_{\geq x}) - v(\pi_{> x})\right)$ for any $v \in G^X$ and $x \in X$,
    where $\Pi_X$ is the set of permutations of $X$ and $\pi_{\triangleright x} \coloneqq \{y \in X \mid \pi(y) \triangleright \pi(x) \}$ for $\triangleright \in \{\geq, >\}$.
\end{lem}

\emph{Games.} A \emph{game arena} is a tuple $\ltuple \sSafe, \sReach, \gameTrans, \sInit \rtuple$ where $\sSafe$ is the set of states controlled by $\pSafe$, $\sReach$ is the set of states controlled by $\pReach$ (and we write $\sAll \coloneqq \sSafe \dot{\cup} \sReach$), $\gameTrans$ is the transition relation on $\sAll$ and $\sInit \in \sAll$ is the initial state.

A \emph{game} consists of a game arena $\ltuple \sSafe, \sReach, \gameTrans, \sInit \rtuple$ and a \emph{winning condition} $\Omega \subseteq \sAll^\omega$. A \emph{play} $\rho \in \sAll^\omega$ is an infinite sequence $\rho_0 \rho_1 \ldots$ such that $\rho_0 = \sInit$ and $(\rho_i, \rho_{i+1}) \in \, \gameTrans$ for all $i \in \mathbb N$. A play $\rho$ is \emph{winning} for $\pSafe$ if $\rho \in \Omega$, otherwise it is winning for $\pReach$. A strategy for $\pSafe$ is a function $\sigma \colon \sSafe \gameTrans \sAll$ with $(s, \sigma(s)) \in \, \gameTrans$ for all $s \in \sSafe$ (and strategies for $\pReach$ are defined similarly). A pair of strategies $\ltuple \sigma_{\pSafe}, \sigma_{\pReach} \rtuple$ for $\pSafe$ and $\pReach$ induces a play $\rho = \rho_0 \rho_1 \ldots$ with $\rho_{i+1} = \sigma_{\pSafe}(\rho_i)$ if $\rho_i \in \sSafe$ and $\rho_{i+1} = \sigma_{\pReach}(\rho_i)$ otherwise. A strategy for $\pSafe$ is winning if, for all strategies of $\pReach$, the induced play is winning for $\pSafe$ (and winning strategies for $\pReach$ are defined similarly).

Given $\Sbad \subseteq \sAll$, a safety winning condition has the form $\Omega_{\Sbad} = \{ \rho \mid \forall i \in \mathbb{N}\colon \rho_i \not\in \Sbad \}$. A safety game consists of a game arena and $\Sbad \subseteq \sAll$ and we write $\ltuple \sSafe, \sReach, \gameTrans, \sInit, \Sbad \rtuple$ instead of $\ltuple \ltuple \sSafe, \sReach, \gameTrans, \sInit \rtuple, \Omega_{\Sbad} \rtuple$.

\section{Optimistic and Pessimistic Responsibility}
\label{sec:responsibility}

To use \semivaluesName{}, we must formulate the responsibility problem as a cooperative game. Therefore, we need to define a value function which takes a coalition of states $C$ and returns $0$ or $1$ depending on whether $C$ can avoid $\Sbad$.

First, we define a safety game where the run of the counterexample is \emph{engraved}: For every state that is on the run and not in $C$, we remove the outgoing transitions except for the transition that follows the run.

\begin{defin}[From transition systems to safety games]
    Let $\ts = \tsTuple$ be a transition system, $\ce=\ce_0 \ldots \ce_k$ a counterexample and $\coalition \subseteq \sAll$. We define $\tsGame{\coalition}= \ltuple \coalition, \sAll \setminus \coalition, \gameTrans', \sInit, \Sbad \rtuple$, where 
    $\ce_i \gameTrans' \ce_{i+1}$ for $i \in \{0, \ldots, k - 1\}$ if $\ce_i \not\in \coalition$ and
    $s \gameTrans' s'$ if $s \tsTrans s'$ and if $s \notin \ce$ or $s \in C$.
    
\end{defin}

We now use this to define a cooperative game.

\begin{defin}[Optimistic and pessimistic cooperative games] Let $\ts = \tsTuple$ be a transition system, $\ce$ a counterexample and $\coalition \subseteq \sAll$. Then the optimistic cooperative game $\voptfull$ is defined as
    \begin{align*}
        \voptfull(\coalition) &= \left\{
        \begin{array}{ll}
            1 & \mbox{if player $\pSafe$ wins } \tsGame{\coalitionOpt} \\
            0 & \mbox{otherwise}
        \end{array}
        \right. \\
        \intertext{and the pessimistic cooperative game $\vpesfull$ is defined as}
        \vpesfull(\coalition) &= \left\{
        \begin{array}{ll}
            1 & \mbox{if player $\pSafe$ wins } \tsGame{C} \\
            0 & \mbox{otherwise.}
        \end{array}
        \right.
    \end{align*}
\end{defin}

In the following, let $\ts = \tsTuple$ be a transition system and $\ce = \ce_0 \ldots \ce_k$ a counterexample. We set $n \coloneqq \left| \tsStates \right|$. We write $\vopt$ instead of $\voptfull$ and $\vpes$ instead of $\vpesfull$.

\begin{ex}
    Consider the railway network from Figure~\ref{fig:train_example} with $\Sbad = \{ \text{\xmark} \}$ and $\rho = s_1 s_2 \text{\xmark}$. Let $C = \varnothing$ (and thus $\coalitionOpt = \{ s_3, \text{\cmark} \}$). In both $\tsGame{C}$ and $\tsGame{\coalitionOpt}$, the only transition from $s_1$ goes to $s_2$ and the only transition from $s_2$ goes to \xmark{}. Therefore, $\pSafe$ cannot win and thus $\vopt(C) = \vpes(C) = 0$.

    Now let $C' = \{ s_1 \}$ and therefore $\coalitionOpt[C'] = \{ s_1, s_3, \text{\cmark}\}$. Then $\pSafe$ wins $\tsGame{\coalitionOpt[C']}$ by playing $s_1 s_3 \text{\cmark} ^\omega$, which are all states controlled by her. However, she loses $\tsGame{C'}$, because either $s_2$ or $s_3$ is reached. $\pReach$ can then move to \xmark{}. Therefore, $\vopt(C') = 1$ and $\vpes(C') = 0$.
\end{ex}

\begin{prop}
    Let $\coalition \subseteq \tsStates$ and $s \in \tsStates$. Then we have $\vopt(\coalition) \geq \vpes(\coalition)$.
\end{prop}

\begin{proof}
If $\vpes(C) = 1$, then $\pSafe{}$ wins $\tsGame{C}$. Because reachability games are monotonic, $\pSafe{}$ also wins $\tsGame{C \cup (S \setminus \ce)}$ and thus $\vopt(C) = 1$. If $\vpes(C) = 0$, then the inequality holds for every value of $\vopt(C)$.
\end{proof}

The forward view, as presented by \citet{ForwardResponsibility}, is closely related to our pessimistic definition. This is because the optimistic definition does not work without a counterexample. Every state in $C$ would belong to \pSafe{}, but every state not in $C$ would also be given to \pSafe{} under the optimistic assumption, so the outcome of the game would not be affected by the value of $C$.%

\begin{defin}[Responsibility]
    Let $\semivalue$ be a \semivalueName{} on $G^S$.
    The \textit{optimistic responsibility of state $s$ with respect to $\semivalue$} is $\semivalue(\vopt, s)$ and
    the \textit{pessimistic responsibility of state $s$ with respect to $\semivalue$} is $\semivalue(\vpes, s)$.
\end{defin}

For example, the pessimistic responsibility of $s$ with respect to the Banzhaf value is $\ban(\vpes, s)$.

\begin{ex}
    We omit \cmark{} and \xmark{} from the coalitions, as control over them never affects the outcome. In the pessimistic case, $\criticalPair{C}{s_2}$ forms a critical pair for $C \in \{ \varnothing, \{s_1\}, \{s_3\} \}$. Therefore, the Shapley responsibility of $s_2$ is $\shap(\vpes, s_2) =1 \cdot \weights_0(\shap) + 2 \cdot \weights_1(\shap) = \frac{2}{3}$. For $s_1$, only $\criticalPair{\{s_3\}}{s_1}$ forms a critical pair and for $s_3$, only $\criticalPair{\{s_1\}}{s_3}$ forms a critical pair, so we have $\shap(\vpes, s_1) = \shap(\vpes, s_3) = 1 \cdot \weights_1(\shap) = \frac{1}{6}$. The full responsibility values for the optimistic and pessimistic case are given in Table~\ref{tab:example_resps}. While it may seem counterintuitive that $s_3$ has positive responsibility even though it was not involved in reaching \xmark, this makes sense upon closer inspection: $s_1$ can only avoid \xmark{} if $s_3$ helps, so it is natural that they share the responsibility.
\end{ex}

\begin{table}[H]
    \centering
	\begin{tabular}{c|ccccc}
		State $s$ & $s_1$ & $s_2$ & $s_3$ & \cmark & \xmark \\
		\hline
		$\shap(\vpes, s)$ & $1/6$ & $2/3$ & $1/6$ & $0$ & $0$ \\
		$\shap(\vopt, s)$ & $1/2$ & $1/2$ & $0$ & $0$ & $0$
	\end{tabular}
   \caption{Optimistic and pessimistic Shapley responsibilities for the train example from Figure~\ref{fig:train_example}.}
    \label{tab:example_resps}
\end{table}

Note that the sum of the responsibility of $s_1$, $s_2$ and $s_3$ is~$1$. This is a general property of the Shapley value (but not of other \semivaluesName{}, such as the Banzhaf value).

\begin{prop} If there exists a path in $\ts$ which does not reach $\Sbad$, then we have
    \[\sum_{s \in S} \shap(\vopt, s) = \sum_{s \in S} \shap(\vpes, s) = 1.\]

\end{prop}

\begin{proof}
Let $\coopGame \in \{\vopt, \vpes\}$. By Lemma~\ref{eqDefShapley}, we have
$\shap(\coopGame, s) = \sum_{\pi \in \Pi_S}\left(\coopGame(\pi_{\geq s}) - \coopGame(\pi_{>s})\right)$.
Because $\ce$ is a counterexample, we have $\coopGame(\varnothing) = 0$ and because there exists a path that does not reach $\Sbad$, we have $\coopGame(S) = 1$. As $\coopGame$ is monotonic, for each permutation $\pi$, there is exactly one state $s$ with $\coopGame(\pi_{\geq s}) - \coopGame(\pi_{>s}) = 1$. We call this state $\switch(\pi)$. For all other states, the difference is $0$. Therefore, we have
\begin{align*}
    \sum_{s\in S} \shap (\coopGame,s) = &  \sum_{s\in S} \frac{1}{n!} \sum_{\pi \in \Pi_S} \left( \coopGame(\pi_{\geq s}) - \coopGame(\pi_{> s}) \right)&  \\
    = & \frac{1}{n!} \sum_{\pi \in \Pi_S} \sum_{s\in S} \left( \coopGame(\pi_{\geq s}) - \coopGame(\pi_{> s}) \right) &\\
    = & \frac{1}{n!} \sum_{\pi \in \Pi_S} \left( \coopGame(\pi_{\geq \switch(\pi)}) - \coopGame(\pi_{> \switch(\pi)}) \right) &\\
    = & \frac{1}{n!} \sum_{\pi \in \Pi_S} 1 = 1. & \qedhere
\end{align*}
\end{proof}

States that only have a single outgoing transition always have responsibility $0$. On the other hand, there can be states with multiple outgoing transitions that have responsibility $0$, e.g. if all outgoing paths from a state eventually reach $\Sbad$.

\section{Algorithms}
\label{sec:complexity}

We investigate the complexity of two decision problems and one counting problem. Each problem has an optimistic variant (where $\vfull = \voptfull$) and a pessimistic variant (where $\vfull = \vpesfull$). A \semivalueName{} $\semivalue$ is encoded by encoding $\weights_i(\semivalue)$ for $i \in \{0, \ldots, n - 1 \}$.

\vspace{0.5em}
\noindent \textbf{Positivity Problem}
\[
\begin{cases}
	\text{Input: }& \ts = \tsTuple, \text{ counterexample }\rho, \\
    & s \in \sAll, \text{\semivalueName{} } \semivalue\\
	\text{Output: }& \text{Is } \semivalue(\vfull, s) > 0?
\end{cases}
\]
\noindent \textbf{Threshold Problem}
\[
\begin{cases}
	\text{Input: }& \ts = \tsTuple, \text{ counterexample }\rho, \\
    & s \in \sAll, \text{\semivalueName{} } \semivalue, t \in [0,1)\\
	\text{Output: }& \text{Is } \semivalue(\vfull, s) > t?
\end{cases}
\]
\noindent \textbf{Computation Problem}
\[
\begin{cases}
	\text{Input: }& \ts = \tsTuple, \text{ counterexample }\rho, \\
    & s \in \sAll, \text{\semivalueName{} } \semivalue\\
	\text{Output: }& \text{What is the value of } \semivalue(\vfull, s)?
\end{cases}
\]

\subsection{Optimistic Responsibility}

We show that the optimistic responsibility is characterised by a simple function. This characterisation then yields an efficient algorithm for computing optimistic responsibility.

For the characterisation, we first define the sets of winning and responsible states.

\begin{defin}[$\ws, \rs{\semivalue}$] Let $\semivalue$ be a \semivalueName{}. The set of \textit{winning states} is $\ws = \{s \in S \mid \vopt(\{s\})=1 \}$, i.e. the set of the states that can win on their own. The set of \textit{responsible states} is $\rs{\semivalue} = \{s \in S \mid \semivalue(\vopt, s)>0 \}$, i.e. the set of the states that have a strictly positive responsibility.
\end{defin}

The following lemma  relates states that can win on their own and states with positive responsibility.

\begin{lemrep}
	\label{lem:rs_in_ws}
	Let $\semivalue$ be a \semivalueName{}. For any $s \in \sAll$, we have
	$\semivalue(\vopt, s) > 0 \implies \vopt (\{s\}) = 1$, i.e. $\rs{\semivalue} \subseteq \ws$.
	\label{thm:pos_responsibility_if_winning_alone}
	
	If $\weights_0(\semivalue) > 0$, the converse implication 
	$\semivalue(\vopt, s) > 0 \impliedby \vopt (\{s\}) = 1$ also holds
	and thus $\rs{\semivalue} = \ws.$
\end{lemrep}

\begin{appendixproof}
	$\implies$: 
	If $\semivalue(\vopt, s) > 0$, there exists a coalition $\coalition \subseteq \tsStates$ with $\vopt(\coalition) = 0$ and $\vopt(\coalition \cup \{s\}) = 1$ (note that $s \not\in \coalition$). Then $\pSafe$ wins $\tsGame{\coalition \cup \{s\} \cup (\tsStates \setminus \Srho)}$ and in particular, $s$ is in \pSafe{}'s winning region. If $s$ were not in \pSafe's winning region, then \pSafe{} could force a win without visiting $s$ and thus would also have a winning strategy in $\tsGame{\coalition \cup (\tsStates \setminus \Srho)}$ (but $\vopt(\coalition) = 0$).
	
	Let $i \in \{1, \ldots, k \}$ such that $s = \ce_i$. This is well-defined as $\ce$ is loop-free by definition. Then, for any $j \in \{1, \ldots, k \}$, $\ce_j$ is in the winning region in $\tsGame{\coalition \cup \{s\} \cup (\tsStates \setminus \Srho)}$ if and only if $j \leq i$: If a state $\ce_j$ with $j > i$ were in the winning region, then \pSafe{} would win $\tsGame{\coalition \cup (\tsStates \setminus \Srho)}$, which contradicts $\vopt(\coalition) = 0$. Similarly, for each $\ce_j$ with $j < i$, \pSafe{} can ensure $\ce_{j+1}$ is reached in the next step (either this is the only option or \pSafe{} controls $\ce_j$). Therefore, \pSafe{} can ensure $s$ is reached from every $\ce_j$ with $j < i$ and thus each such $\ce_j$ is in the winning region.
	
	We now set $\coalition' = \coalition \setminus \{\ce_j \in \ce \mid j > i \}$, define $\tsGameSymb' \coloneqq \tsGame{\coalition' \cup \{s\} \cup (\tsStates \setminus \ce)}$ and show that \pSafe{} wins $\tsGameSymb'$. As no $\ce_j$ with $j > i$ is in the winning region, \pSafe{} has a winning strategy that ensures no such $\ce_j$ is reached. Therefore, this is also a winning strategy in $\tsGameSymb'$.
	
	We now set $\coalition'' = \coalition' \setminus \{\ce_j \in \ce \mid j < i \}$, define $\tsGameSymb'' \coloneqq \tsGame{\coalition'' \cup \{s\} \cup (\tsStates \setminus \ce)}$ and show that \pSafe{} wins $\tsGameSymb''$. For this, let $\sigma$ be a winning strategy for \pSafe{} in $\tsGameSymb'$. For any $j < i$, the only transition in $\tsGameSymb''$ from $\ce_j$ is to $\ce_{j+1}$. Therefore, any play that reaches $\ce_j$ eventually reaches $s$. Therefore, $\sigma$ (restricted to $\coalition'' \cup \{s\} \cup (\tsStates \setminus \ce)$) is a winning strategy for \pSafe{} in $\tsGameSymb''$, as any play is either unchanged or eventually reaches $s$ and thus forms a loop.
	
	Note that $\coalition'' \cap \Srho = \varnothing$, because $\rho_i = s \not\in \coalition$ and all other $\ce_j \in \rho$ have been excluded. Therefore, we have $\tsGame{\coalition'' \cup \{s\} \cup (\tsStates \setminus \ce)} = \tsGame{\{s\} \cup (\tsStates \setminus \ce)}$ and thus $\vopt(\{s\}) = 1$.\\

	$\impliedby$:We have 
	\[\semivalue(\vopt, s) = \!\!\!\!\!\! \sum_{C \subseteq \sAll \setminus \{s\}} \!\!\!\!\! \weights_{|C|}(\semivalue)[\vopt(C\cup \{s\})-\vopt(C)].\]
	
	As $\vopt(\varnothing) = 0$ always holds, $\weights_0(\semivalue)(\vopt(\{s\}) - \vopt(\varnothing)) > 0$. Furthermore, $\vopt$ is monotone, so no term in the sum is negative. Therefore, $\semivalue(\vopt, s) > 0$.
\end{appendixproof}

\begin{correp}
	If $\weights_i(\semivalue) > 0$ for $i \in \{0, \ldots, n - 1\}$, then \(\semivalue(\vopt, s) > 0 \implies \semivalue(\vpes, s) > 0.\)
\end{correp}

\begin{appendixproof}
	If $\weightVec_0 > 0$, we can apply Lemma~\ref{lem:rs_in_ws} to deduce that $\vopt(\{s\}) = 1$, and $\vopt(\varnothing)=0$ always holds, so $\weightVec_1(\vopt(\{s\}) - \vopt(\varnothing))= \weightVec_1 > 0 $ and thus $\semivalue(\vpes, s) > 0$.
\end{appendixproof}

Using Lemma~\ref{lem:rs_in_ws}, the following theorem then shows that responsibility in the optimistic case is a yes-or-no question.
\begin{thm}[Characterisation of optimistic responsibility]
	\label{thm:optimistic_constant}
	Let $\semivalue$ be a \semivalueName{}, then we have
	\[ \semivalue(\vopt, s) = \begin{cases}
		K & \text{if } s \in \ws \\
		0 & \text{otherwise}
	\end{cases} \]
	with $K = \displaystyle \sum_{i=0}^{n-w} \binom{n-w}{i} \cdot \weights_i(\semivalue) $ and $w \coloneqq |\ws|$.
	Additionally, we have $\ws \subseteq \Srho$.      
\end{thm}
The relevant question in the optimistic case is therefore ``Is $s$ in $\ws$?'', i.e. ``Can $s$ win on its own?''.

\begin{proof}
	Let $s \in \ws$ and let $\weightVec = \weightVecTuple{n-1}$ with $\weightVec_i = \weights_i(\semivalue)$. Then \[\semivalue(\vopt, s) = \sum_{C \subseteq S \setminus \{s\}} p_{|C|}[\vopt(C\cup \{s\}) - \vopt(C)].\]
	
	Let $C \subseteq S \setminus \{s\}$. Lemma~\ref{lem:rs_in_ws} implies that $\vopt(C\cup \{s\}) - \vopt(C) = 1$ if and only if $C \cap \ws = \varnothing$ and $\vopt(C\cup \{s\}) - \vopt(C) = 0$ otherwise.
	In the former case, $C \subseteq S \setminus \ws$ and $\left|S \setminus \ws\right| = n - w$. Therefore, there are $\binom{n-w}{i}$ coalitions of size $i$ that fulfil the condition and we have
	\begin{align*}
		\semivalue(\vopt, s) & = \sum_{C \subset S \setminus \ws} p_{|C|}[\vopt(C\cup \{s\}) - \vopt(C)] \\
		& = \sum_{C \subset S \setminus \ws} p_{|C|} 
		 = \sum_{i=0}^{n-w} \binom{n-w}{i} p_i.
	\end{align*}    
	
	We now show that $\ws \subseteq \Srho$. By Lemma~\ref{lem:rs_in_ws}, if $\weights_0(\semivalue) > 0$ and $\semivalue(\vopt, s) > 0$, then $s$ can avoid reaching $\Sbad$ on its own. If $s \not\in \Srho$, this is impossible.
\end{proof}

In the case of the Shapley and Banzhaf values, we can give a closed form for the constant $K$:
\begin{proprep}
	\label{prop:optimistic_shapley_constant_K}
	For any $s \in \ws$, we have
	\begin{center}
		$\shap(\vopt, s) = \displaystyle \frac{1}{|\ws|}$ \hspace{1em} and \hspace{1em}
		$\ban(\vopt, s) = \displaystyle \frac{1}{2^{|\ws|-1}}.$
	\end{center}
\end{proprep}

\begin{appendixproof}
	Recall the alternative definition of the Shapley value from Lemma~\ref{eqDefShapley}:
	\[\shap(\vopt, s) = \frac{1}{n!} \sum_{\pi \in \Pi_S} \vopt(\pi_{\geq s}) - \vopt(\pi_{> s}).\]
	We count the number of permutations $\pi$ such that $\pi_{\geq s} \cap \ws = \{s\}$. As described in the proof of Theorem~\ref{thm:optimistic_constant}, only these $\pi$ give positive terms in the sum.
	
	Multiple permutations may produce the same coalition: Let $C \coloneqq \pi_{\geq s}$ with $\left|C\right| = i$. Then there are $(n-i)! (i-1)!$ different permutations $\pi'$ such that $\pi'_{\geq s} = C$, as there are $n - i$ elements before $s$ and $i - 1$ elements after $s$ and the order of both groups is irrelevant.
	
	The number of coalitions $C$ of size $i$ such that $C \cap \ws = \{s\}$ is     
	\begin{align*}
		& |\{ C \subseteq S \mid C \cap \ws = \{s\}, \left|C\right|=i \}| \\
		= & |\{ C \cup \{s\} \mid C \subseteq S, C \cap \ws = \emptyset, |C|=i-1 \}| \\
		= & \binom{n-w}{i-1}.
	\end{align*}
	
	Therefore, for each $i \in \{1, \ldots, n - w + 1\}$, there are $\binom{n-w}{i-1}$ coalitions of size $i$ with $ \vopt(C) - \vopt(C \setminus \{s\}) = 1$. Each of these coalitions is produced by $(n-i)!(i-1)!$ different permutations (and each permutation only produces one coalition). Therefore, the Shapley value is given by:
	
	\begin{align*}
		\shap(\vopt, s) & = \frac{1}{n!} \sum_{i=1}^{n-w} \binom{n-w}{i-1} (n-i)!(i-1)! \\
		& = \frac{1}{n!} \sum_{i=w}^{n} \binom{n-w}{i-w} (n-i+w-1)!(i-w)! \\
		& = \sum_{i=w}^{n} \frac{(n-w)!(n-i+w-1)!(i-w)!w!}{(n-i)!(i-w)!n!w!} \\
		& = \frac{1}{w \binom{n}{w}} \sum_{i=w}^{n} \frac{(n-i+w-1)!}{(n-i)!(w-1)!} \\
		& = \frac{1}{w \binom{n}{w}} \sum_{i=w}^{n} \binom{n-i}{w-1}.
	\end{align*}
	
	We prove by induction that, for all $n \geq w$, we have \[\binom{n}{w} = \sum_{i=1}^{n-w+1} \binom{n-i}{w-1}.\]
	
	For $n = w$, we get
	\[\binom{w}{w} = \binom{w-1}{w-1} = \sum_{i=1}^{1} \binom{w-i}{w-1}.\]
	
	Now suppose
	\[\displaystyle \binom{n-1}{w} = \sum_{i=1}^{n-1-w+1} \binom{n-1-i}{w-1}.\]
	Then we have    
	\begin{align*}
		\binom{n}{w} = & \displaystyle \binom{n-1}{w} + \binom{n-1}{w-1} \\
		= & \left( \sum_{i=1}^{n-1-w+1} \binom{n-1-i}{w-1}\right) + \binom{n-1}{w-1}\\
		= & \left( \sum_{i=2}^{n-1-r+2} \binom{n-i}{w-1} \right) + \binom{n-1}{w-1} \\
		= & \sum_{i=1}^{n-1-w+2} \binom{n-1-i}{w-1}.
	\end{align*}
	
	Therefore, we can conclude that $\shap(\vopt, s) = \frac{1}{w}$.\\
	
	In the case of Banzhaf, by Theorem~\ref{thm:optimistic_constant}, we get \[\ban(\vopt,s) = \sum_{i=0}^{n-w} \binom{n-w}{i} \weights_i(\ban)\] and for all $i \in \{0, \ldots, n-w \}$, $\weights_i(\ban) = \frac{1}{2^{n-1}}$. Therefore, we have
	\begin{align*}
		\ban(\vopt,s) = & \sum_{i=0}^{n-w} \binom{n-w}{i} \cdot \frac{1}{2^{n-1}} \\
		= & \frac{1}{2^{n-1}} \cdot \sum_{i=0}^{n-w} \binom{n-w}{i} \\
		= & \frac{1}{2^{n-1}} \cdot 2^{n-w} \\
		= & \frac{1}{2^{w - 1}}.
	\end{align*}

\end{appendixproof}

Table~\ref{tab:example_resps} shows that pessimistic responsibility does not satisfy this property, as $s_1$ and $s_2$ have different positive responsibilities.

This characterisation now yields an efficient algorithm for optimistic responsibility. We compute the set of winning states $\ws$ as follows. For each state $s \in \sAll$, we determine whether $\{s \}$ is a winning coalition by constructing $\tsGame{\{s\}}$ and solving it with the attractor algorithm \cite{DBLP:conf/dagstuhl/2001automata}, which has linear runtime. We invoke it $\left| S \right|$ times, yielding quadratic runtime. Due to Lemma~\ref{lem:rs_in_ws}, if $s \not\in \ws$, the responsibility of $s$ is $0$. Otherwise, we can compute the responsibility constant $K$ from Theorem~\ref{thm:optimistic_constant} and return it. This yields the following theorem.
\begin{thm}
    The optimistic positivity, threshold and computation problems are solvable in polynomial time.
\end{thm}

\subsection{Pessimistic Responsibility}

Computing pessimistic responsibility is more difficult:

\begin{proprep}
\label{prop:pessimistic_positivity_np_complete}
    The pessimistic positivity problem is NP-complete.
\end{proprep}

\begin{proofsketch}
	To show inclusion in NP, we nondeterministically guess a coalition $C$ and then verify that $\weights_{\left|C\right|}(\semivalue) (\vpes(C \cup \{s\}) -\vpes(C)) > 0$. This requires polynomial time and thus, the problem is in NP.
	
	To show NP-hardness, we give a reduction from the forward responsibility positivity problem, which \citet{ForwardResponsibility} have shown to be NP-hard. For this, we take the given transition system $\ts$ and and construct a new transition system $\ts'$ with new initial state $s'_0$ and edges from $s'_0$ to $s_0$ and $\Sbad$. We choose $\ce = s'_0 \bar{s}$ for some $\bar{s} \in \Sbad$. Then a coalition $C$ is winning in $\ts$ if and only if $C \cup \{s'_0\}$ is winning in $\ts'$ and thus, a state has positive responsibility in $\ts$ if and only if it has positive responsibility in $\ts'$.
\end{proofsketch}

\begin{appendixproof}
    We first show that the positivity problem is in NP. A state $s$ has positive responsibility if there exists a critical pair $\criticalPair{C}{s}$ such that $\weights_{|C|}(\semivalue) > 0$. We non-deterministically guess $C$, verify that $\weights_{|C|}(\semivalue) > 0$ and check whether $\criticalPair{C}{s}$ is a critical pair. For this, we solve the games $\mathcal{G} = \tsGame{C}$ and $\mathcal{G}' =\tsGame{C \cup \{s\}}$ and accept if and only if $\pSafe$ loses $\mathcal{G}$ and wins $\mathcal{G'}$.    
    
    The \emph{forward responsibility positivity problem} is defined as follows:
    \[
    \begin{cases}
    	\text{Input: }& \ts = \tsTuple, s \in \sAll\\
    	\text{Output: }& \text{Is the responsibility of }s\text{ positive?}
    \end{cases}
    \]
    
    \citet{ForwardResponsibility} show that it is NP-hard. We present a reduction from the forward responsibility positivity problem to the backward responsibility positivity problem.
    
    Let $\ts, s$ with $\ts = \tsTuple$ and $s \in \sAll$ be an instance of the forward positivity computation problem with $\left| \sAll \right| = n$. If $\Sbad = \varnothing$, then all states have responsibility $0$. We output an instance of the backward positivity problem where all states have responsibility $0$. We can construct such an instance by constructing a transition system $\ts' =(\{s_0, \ldots, s_n\}, \tsTrans', s_0, \{s_n\})$, where $s_i \tsTrans' s_{i+1}$ for $i \in \{0, \ldots, n-1\}$. We set $\rho = s_0 \ldots s_n$. As every state only has one successor, they all have responsibility $0$.
    
    If $\Sbad \neq \varnothing$, then let $\bar{s} \in \Sbad$. We construct $\ts' = \ltuple \sAll', \tsTuple \cup \{(s'_0, s_0), (s'_0, \bar{s})\}, \Sbad \rtuple$ with $\sAll' = \sAll \dot{\cup} \{ s'_0 \}$ and set $\ce = s'_0 \bar{s}$.
    
    We now show that $s$ has positive forward responsibility in $\ts$ if it has positive backward responsibility in $\ts'$. If $s$ has positive backward responsibility in $\ts'$, then there exists a critical pair $\criticalPair{C}{s}$ in $\ts'$. Therefore, $\pSafe$ has a winning strategy in $\tsGame{C}$. Because the only two transitions from $s'_0$ go to $\sbar \in \Sbad$ and $s_0$, the strategy of $\pSafe$ must choose $s_0$. Therefore, $\pSafe$ also wins the game corresponding to $\ts$ and $C \setminus \{s'_0\}$ with the same strategy. Therefore, $s$ has positive forward responsibility in $\ts$.

    Therefore, the backward responsibility problem is NP-complete.
\end{appendixproof}

\begin{proprep}
    \label{prop:pes_threshold_problem_complexity}
    The pessimistic threshold problem is NP-hard and in PSPACE.
\end{proprep}

\begin{proofsketch}
	NP-hardness follows from the NP-hardness of the positivity problem shown in Proposition~\ref{prop:pessimistic_positivity_np_complete} by choosing threshold $t=0$.
	
	To show inclusion in PSPACE, we construct an algorithm that computes the responsibility in polynomial space. For this, we count the number of coalitions of each size that are significant, multiply the counts by the corresponding weights, and then output the sum.
\end{proofsketch}

\begin{appendixproof}
	NP-hardness follows from the NP-hardness of the positivity problem shown in Proposition~\ref{prop:pessimistic_positivity_np_complete}. We can reduce the positivity problem to the threshold problem by choosing threshold $t = 0$.
	
    To show inclusion in PSPACE, let $\ts = \tsTuple$, $s \in \sAll$, $\semivalue$, $\Sbad \subseteq \sAll$, $\rho$, $t \in [0,1)$ be the input of the pessimistic threshold problem.
    
    We first initialise integer counters $c_0, \ldots, c_{n-1}$ with value $0$. We then iterate over all coalitions $C \subseteq \sAll \setminus \{s\}$. If $\criticalPair{C}{s}$ is a critical pair, we increment $c_{|C|}$ by one. Finally, we compute the responsibility of $s$ by computing the sum $\sum_{i = 0}^{n-1} c_i \cdot \weights_i(\semivalue)$.

    This algorithm requires polynomial space as every $c_i$ is bounded by $2^{n}$ and thus requires linear space. Solving a game using the attractor algorithm also only requires polynomial space.
\end{appendixproof}

The pessimistic threshold problem thus lies between NP and PSPACE, but the precise complexity is still open.

\begin{proprep}
    The pessimistic computation problem is \#P-hard and can be solved in polynomial space.
\end{proprep}

\begin{proofsketch}
	A polynomial-space algorithm is given in Proposition~\ref{prop:pes_threshold_problem_complexity}.
    \#P-hardness can be shown by reduction from the forward computation problem \cite{ForwardResponsibility}, which is \#P-complete. We perform the same construction as in Proposition~\ref{prop:pessimistic_positivity_np_complete} and show that there is a bijection between winning coalitions in the forward and in the backward view.
\end{proofsketch}

\begin{appendixproof}
	The \emph{forward responsibility computation problem} is defined as follows:
	\[
	\begin{cases}
		\text{Input: }& \ts = \tsTuple, s \in \sAll\\
		\text{Output: }& \text{What is the forward responsibility of }s\text{?}
	\end{cases}
	\]
	
	\citet{ForwardResponsibility} show that it is \#P-hard. We present a reduction from the forward responsibility computation problem to the (backward) responsibility computation problem.
	
	Let $\ts, s$ with $\ts = \tsTuple$ and $s \in \sAll$ be an instance of the forward responsibility computation problem with $\left| \sAll \right| = n$. If $\Sbad = \varnothing$, then all states have responsibility $0$. We output an instance of the backward responsibility problem where all states have responsibility $0$. Such an instance always exists, as shown in the proof of Proposition~\ref{prop:pessimistic_positivity_np_complete}.
	
	Otherwise, $\Sbad \neq \varnothing$. Let $\bar{s} \in \Sbad$. We construct $\ts' = \ltuple \sAll', \tsTuple \cup \{(s'_0, s_0), (s'_0, \bar{s})\}, \Sbad \rtuple$ with $\sAll' = \sAll \dot{\cup} \{ s'_0 \}$ and set $\ce = s'_0 \bar{s}$. We define the \semivalueName{} $\shap'$ with $\weights_0(\shap') = 0$ and $\weights_i(\shap') = \weights_{i-1}(\shap)$ for $i \in \{1, \ldots, n\}$. The \semivalueName{} $\shap'$ is thus a shifted version of the Shapley value. We now output the backward responsibility computation problem instance $\ts', \ce, s$ and $\shap'$.
	
	The value computed from this instance is equal to the value computed for the forward computation problem instance $\ts, s$:
	\newcommand{\vForward}{v_{\mathit{fw}}}
	\newcommand{\vBackward}{v_{\mathit{bw}}}
	
	Let $\vForward$ be the cooperative game in the forward case, i.e. $\vForward(C) = 1$ if \pSafe{} can win in the forward safety game by controlling $C$. Analogously, we define $\vBackward$ as the cooperative game in the backward case. We first show that, for all $C \subseteq \sAll$, $\vForward(C) = 1$ if and only if $\vBackward(C \cup \{s'_0\}) = 1$:
	
	Let $\vForward(C) = 1$. Then there is a winning strategy for \pSafe{} that avoids $\Sbad$ from $s_0$. Then $\vBackward(C \cup \{s'_0\}) = 1$, as \pSafe{} can first go from $s'_0$ to $s_0$ and then follow the same strategy as in the forward case.
	
	Now let $\vBackward(C \cup \{s'_0\}) = 1$. Then there is a winning strategy for \pSafe{} that avoid s$\Sbad$ from $s'_0$. As $s'_0$ only has two transitions -- one to $s_0$ and one to $\bar{s} \in \Sbad$ -- \pSafe{} must go to $s_0$ from $s'_0$. After that, the play remains in $\sAll$, as $s'_0$ is unreachable from $s_0$. Furthermore, the counterexample does not restrict the behaviour of any $s \in \sAll$. Therefore, \pSafe{} also wins from $s_0$ in the forward case and thus $\vForward(C) = 1$.
	
	Furthermore, $\vBackward(C) = 0$ for all $C \subseteq \sAll$ with $s'_0 \notin C$.
	
	The forward responsibility of $s$ is then given by the following sum and we have
	
	\newcommand{\negSpaceBefore}{\!\!\!\!}
	\newcommand{\negSpaceAfter}{\!\!\!}
	\begin{align*}
		& \negSpaceBefore{}\sum_{C \subseteq \sAll \setminus \{s\}} \negSpaceAfter{}\weights_{\left|C \right|}(\shap)  \cdot \left( \vForward(C \cup \{ s\}) - \vBackward(C) \right) \\
		= & \negSpaceBefore{}\sum_{C \subseteq \sAll \setminus \{s\}} \negSpaceAfter{}\weights_{\left|C\right|}(\shap) \\
		& \qquad \qquad \cdot ( \vForward(C \cup \{s'_0, s\}) - \vBackward(C \cup \{s'_0\}) ) \\
		= & \negSpaceBefore{}\sum_{C \subseteq \sAll \setminus \{s\}} \negSpaceAfter{}\weights_{\left|C \cup \{ s'_0 \}\right|}(\shap')  \\
		& \qquad \qquad \cdot (\vForward(C \cup \{s'_0, s\}) - \vBackward(C \cup \{s'_0\}) ) \\
		= & \negSpaceBefore{}\sum_{C' \subseteq \sAll' \setminus \{s\}} \negSpaceAfter{}\!\!\! \weights_{\left|C'\right|}(\shap') \left( \vForward(C' \cup \{s\}) - \vBackward(C') \right) \\
		= & \enspace \semivalue(\vBackward, s),
	\end{align*}
	which is the value of the backward responsibility.
\end{appendixproof}

\section{Implementation}
\label{sec:implementation}

We have developed a tool that computes optimistic and pessimistic backward responsibility\footnote{Available at https://github.com/johannesalehmann/backward-responsibility}
 \cite{lehmann_2024_10610862}. Our tool implements three techniques for this: It supports exact computation of responsibility, which works well on small models, but has exponential runtime. It can also use a stochastic algorithm that produces a good approximation of responsibility for larger models with thousands of states. Finally, it is possible to group states, which makes results less cluttered and improves runtime significantly.

All benchmarks were run on a MacBook Pro running macOS 13.3 with an 8-core M2 chip and 24 GB of memory. The experiments were conducted using the Shapley value, as it is the most widely-used \semivalueName{}.

\subsection{Exact Algorithm}
\newcommand{\Prism}{\textsc{Prism}}

To compute responsibility exactly, our tool uses the model checker \Prism{} \cite{PrismChecker} to build the model and check the safety property. \Prism{} produces a transition system and counterexample as output. Alternatively, this raw model and counterexample can be provided directly by the user. For every coalition and state, we then check whether they form a critical pair. If they do, we increment the result by the value given by the \semivalueName{}.

We found that, in practice, it is faster to first compute the minimal winning coalitions of the game and then analyse the coalitions. This way, each coalition is only solved once. Both steps can be parallelised with minimal overhead.

\subsubsection{Example: Peg Solitaire}

In the single-player board game \emph{Peg Solitaire} \cite{PegSolitaire}, the player is presented with a grid of holes (in our case, the grid is triangle-shaped). All but one of the holes are filled with a peg. In each move, the player may pick any peg and jump over an adjacent peg-filled hole into an empty hole behind that. After that, the peg they jumped over is removed. The goal is to remove all but one of the pegs. It is possible to lose by reaching a configuration with multiple pegs where none of the pegs can jump. Such a configuration is depicted in Figure~\ref{fig:peg_solitaire_losing}.

\newcommand{\hdist}{4.5}
\newcommand{\vdist}{0.86602540378*\hdist}
\newcommand{\hOffset}{0.5*\hdist}
\newcommand{\circPos}[2]{(#1*\hdist-#2*\hOffset, #2*-1*\vdist)}
\newcommand{\filledPeg}[3]{\outerCircle{#1}{#2} \draw[black, fill=black] \circPos{#1}{#2} circle (1.4mm);\pegNumber{#1}{#2}{#3}{white}}
\newcommand{\emptyPeg}[3]{\outerCircle{#1}{#2} \pegNumber{#1}{#2}{#3}{black}}
\newcommand{\outerCircle}[2]{\draw[black, line width=0.5mm] \circPos{#1}{#2} circle (1.9mm);}
\newcommand{\pegNumber}[4]{\node[#4] at \circPos{#1}{#2} {\scriptsize #3};}

\begin{figure}[H]
	\centering
	\begin{subfigure}{3.7cm}
		\centering
		\begin{tikzpicture}[x=1mm,y=1mm]
			\emptyPeg{0}{0}{1}
			\emptyPeg{0}{1}{2}
			\emptyPeg{1}{1}{3}
			\filledPeg{0}{2}{4}
			\emptyPeg{1}{2}{5}
			\emptyPeg{2}{2}{6}
			\emptyPeg{0}{3}{7}
			\emptyPeg{1}{3}{8}
			\emptyPeg{2}{3}{9}
			\emptyPeg{3}{3}{10}
			\filledPeg{0}{4}{11}
			\emptyPeg{1}{4}{12}
			\emptyPeg{2}{4}{13}
			\emptyPeg{3}{4}{14}
			\emptyPeg{4}{4}{15}
		\end{tikzpicture}
		\caption{Final, losing configuration of the game}
		\label{fig:peg_solitaire_losing}
	\end{subfigure}%
	\hspace{1em}
	\begin{subfigure}{3.5cm}
		\centering
		\begin{tikzpicture}[x=1mm,y=1mm]
			\emptyPeg{0}{0}{1}
			\emptyPeg{0}{1}{2}
			\emptyPeg{1}{1}{3}
			\filledPeg{0}{2}{4}
			\filledPeg{1}{2}{5}
			\emptyPeg{2}{2}{6}
			\filledPeg{0}{3}{7}
			\emptyPeg{1}{3}{8}
			\emptyPeg{2}{3}{9}
			\filledPeg{3}{3}{10}
			\filledPeg{0}{4}{11}
			\emptyPeg{1}{4}{12}
			\emptyPeg{2}{4}{13}
			\filledPeg{3}{4}{14}
			\filledPeg{4}{4}{15}
		\end{tikzpicture}
		\caption{Last configuration with positive responsibility}
		\label{fig:peg_solitaire_critical}
	\end{subfigure}
	\caption{Analysis of a Peg Solitaire game.}
	\label{fig:peg_solitaire}
\end{figure}
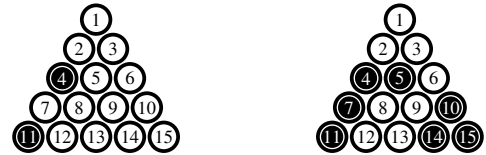

\newcommand{\psMove}{\!\to\!}
Given a play that reached such a configuration, a natural question is to find the last state from which the game was still winnable. One can use optimistic responsibility to determine this. Consider the configuration from Figure~\ref{fig:peg_solitaire_losing}, which was the result starting with Hole $1$ empty and playing $4\psMove1$, $6\psMove4$, $1\psMove6$, $12\psMove5$, $14\psMove12$, $6\psMove13$, $12\psMove14$, $15\psMove13$, $7\psMove2$, $2\psMove9$, $10\psMove8$, $13\psMove4$. A graphical depiction of the full game is given in the appendix. Computing optimistic responsibility reveals that the configuration depicted in Figure~\ref{fig:peg_solitaire_critical} is the last state with positive responsibility. If $15\psMove13$ is played in this state, all further states have responsibility $0$, which tells us that the game is unwinnable. On the other hand, if $7\psMove2$, $2\psMove9$, $15\psMove6$, $6\psMove13$, $14\psMove12$, $11\psMove13$ is played from the configuration in Figure~\ref{fig:peg_solitaire_critical}, the game is won. This demonstrates that optimistic responsibility is useful when only a qualitative analysis is desired.

\begin{toappendix}
	\renewcommand{\hdist}{6}
	\renewcommand{\vdist}{0.86602540378*\hdist}
	\renewcommand{\hOffset}{0.5*\hdist}
	\renewcommand{\circPos}[2]{(#1*\hdist-#2*\hOffset, #2*-1*\vdist)}
	\renewcommand{\filledPeg}[3]{\outerCircle{#1}{#2} \draw[black, fill=black] \circPos{#1}{#2} circle (2.0mm);\pegNumber{#1}{#2}{#3}{white}}
	\renewcommand{\emptyPeg}[3]{\outerCircle{#1}{#2} \pegNumber{#1}{#2}{#3}{black}}
	\renewcommand{\outerCircle}[2]{\draw[black, line width=0.5mm] \circPos{#1}{#2} circle (2.5mm);}
	\renewcommand{\pegNumber}[4]{\node[#4] at \circPos{#1}{#2} {\small #3};}
	\newcommand{\moveDesc}[1]{\node at (0,5) {#1};}
	\newcommand{\hDist}{\hspace{2em}}
	\newcommand{\vDist}{\vspace{0.85em}}
	\subsection{Peg Solitaire Configurations}
	The peg solitaire game from Section~\ref{sec:implementation} had the following configurations:
	
	\begin{tikzpicture}[x=1mm,y=1mm]
		\moveDesc{Initial configuration:}
		\emptyPeg{0}{0}{1}
		\filledPeg{0}{1}{2}
		\filledPeg{1}{1}{3}
		\filledPeg{0}{2}{4}
		\filledPeg{1}{2}{5}
		\filledPeg{2}{2}{6}
		\filledPeg{0}{3}{7}
		\filledPeg{1}{3}{8}
		\filledPeg{2}{3}{9}
		\filledPeg{3}{3}{10}
		\filledPeg{0}{4}{11}
		\filledPeg{1}{4}{12}
		\filledPeg{2}{4}{13}
		\filledPeg{3}{4}{14}
		\filledPeg{4}{4}{15}
	\end{tikzpicture}
	\hDist{}
	\begin{tikzpicture}[x=1mm,y=1mm]
		\moveDesc{After $4\to1$ is played:}
		\filledPeg{0}{0}{1}
		\emptyPeg{0}{1}{2}
		\filledPeg{1}{1}{3}
		\emptyPeg{0}{2}{4}
		\filledPeg{1}{2}{5}
		\filledPeg{2}{2}{6}
		\filledPeg{0}{3}{7}
		\filledPeg{1}{3}{8}
		\filledPeg{2}{3}{9}
		\filledPeg{3}{3}{10}
		\filledPeg{0}{4}{11}
		\filledPeg{1}{4}{12}
		\filledPeg{2}{4}{13}
		\filledPeg{3}{4}{14}
		\filledPeg{4}{4}{15}
	\end{tikzpicture}
	\vDist{}
	
	\begin{tikzpicture}[x=1mm,y=1mm]
		\moveDesc{After $6\to4$ is played:}
		\filledPeg{0}{0}{1}
		\emptyPeg{0}{1}{2}
		\filledPeg{1}{1}{3}
		\filledPeg{0}{2}{4}
		\emptyPeg{1}{2}{5}
		\emptyPeg{2}{2}{6}
		\filledPeg{0}{3}{7}
		\filledPeg{1}{3}{8}
		\filledPeg{2}{3}{9}
		\filledPeg{3}{3}{10}
		\filledPeg{0}{4}{11}
		\filledPeg{1}{4}{12}
		\filledPeg{2}{4}{13}
		\filledPeg{3}{4}{14}
		\filledPeg{4}{4}{15}
	\end{tikzpicture}
	\hDist{}
	\begin{tikzpicture}[x=1mm,y=1mm]
		\moveDesc{After $1\to6$ is played:}
		\emptyPeg{0}{0}{1}
		\emptyPeg{0}{1}{2}
		\emptyPeg{1}{1}{3}
		\filledPeg{0}{2}{4}
		\emptyPeg{1}{2}{5}
		\filledPeg{2}{2}{6}
		\filledPeg{0}{3}{7}
		\filledPeg{1}{3}{8}
		\filledPeg{2}{3}{9}
		\filledPeg{3}{3}{10}
		\filledPeg{0}{4}{11}
		\filledPeg{1}{4}{12}
		\filledPeg{2}{4}{13}
		\filledPeg{3}{4}{14}
		\filledPeg{4}{4}{15}
	\end{tikzpicture}
	\vDist{}
	
	\begin{tikzpicture}[x=1mm,y=1mm]
		\moveDesc{After $12-5$ is played:}
		\emptyPeg{0}{0}{1}
		\emptyPeg{0}{1}{2}
		\emptyPeg{1}{1}{3}
		\filledPeg{0}{2}{4}
		\filledPeg{1}{2}{5}
		\filledPeg{2}{2}{6}
		\filledPeg{0}{3}{7}
		\emptyPeg{1}{3}{8}
		\filledPeg{2}{3}{9}
		\filledPeg{3}{3}{10}
		\filledPeg{0}{4}{11}
		\emptyPeg{1}{4}{12}
		\filledPeg{2}{4}{13}
		\filledPeg{3}{4}{14}
		\filledPeg{4}{4}{15}
	\end{tikzpicture}
	\hDist{}
	\begin{tikzpicture}[x=1mm,y=1mm]
		\moveDesc{After $14\to12$ is played:}
		\emptyPeg{0}{0}{1}
		\emptyPeg{0}{1}{2}
		\emptyPeg{1}{1}{3}
		\filledPeg{0}{2}{4}
		\filledPeg{1}{2}{5}
		\filledPeg{2}{2}{6}
		\filledPeg{0}{3}{7}
		\emptyPeg{1}{3}{8}
		\filledPeg{2}{3}{9}
		\filledPeg{3}{3}{10}
		\filledPeg{0}{4}{11}
		\filledPeg{1}{4}{12}
		\emptyPeg{2}{4}{13}
		\emptyPeg{3}{4}{14}
		\filledPeg{4}{4}{15}
	\end{tikzpicture}
	\vDist{}
	
	\begin{tikzpicture}[x=1mm,y=1mm]
		\moveDesc{After $6\to13$ is played:}
		\emptyPeg{0}{0}{1}
		\emptyPeg{0}{1}{2}
		\emptyPeg{1}{1}{3}
		\filledPeg{0}{2}{4}
		\filledPeg{1}{2}{5}
		\emptyPeg{2}{2}{6}
		\filledPeg{0}{3}{7}
		\emptyPeg{1}{3}{8}
		\emptyPeg{2}{3}{9}
		\filledPeg{3}{3}{10}
		\filledPeg{0}{4}{11}
		\filledPeg{1}{4}{12}
		\filledPeg{2}{4}{13}
		\emptyPeg{3}{4}{14}
		\filledPeg{4}{4}{15}
	\end{tikzpicture}
	\hDist{}
	\begin{tikzpicture}[x=1mm,y=1mm]
		\moveDesc{After $12\!\to\!14$ is played:}
		\emptyPeg{0}{0}{1}
		\emptyPeg{0}{1}{2}
		\emptyPeg{1}{1}{3}
		\filledPeg{0}{2}{4}
		\filledPeg{1}{2}{5}
		\emptyPeg{2}{2}{6}
		\filledPeg{0}{3}{7}
		\emptyPeg{1}{3}{8}
		\emptyPeg{2}{3}{9}
		\filledPeg{3}{3}{10}
		\filledPeg{0}{4}{11}
		\emptyPeg{1}{4}{12}
		\emptyPeg{2}{4}{13}
		\filledPeg{3}{4}{14}
		\filledPeg{4}{4}{15}
	\end{tikzpicture}
	\vDist{}

	This is the last configuration where the game is still winnable. An alternative continuation where the game is won is depicted after this game.
	
	\begin{tikzpicture}[x=1mm,y=1mm]
		\moveDesc{After $15\to13$ is played:}
		\emptyPeg{0}{0}{1}
		\emptyPeg{0}{1}{2}
		\emptyPeg{1}{1}{3}
		\filledPeg{0}{2}{4}
		\filledPeg{1}{2}{5}
		\emptyPeg{2}{2}{6}
		\filledPeg{0}{3}{7}
		\emptyPeg{1}{3}{8}
		\emptyPeg{2}{3}{9}
		\filledPeg{3}{3}{10}
		\filledPeg{0}{4}{11}
		\emptyPeg{1}{4}{12}
		\filledPeg{2}{4}{13}
		\emptyPeg{3}{4}{14}
		\emptyPeg{4}{4}{15}
	\end{tikzpicture}
	\hDist{}
	\begin{tikzpicture}[x=1mm,y=1mm]
		\moveDesc{After $7\to2$ is played:}
		\emptyPeg{0}{0}{1}
		\filledPeg{0}{1}{2}
		\emptyPeg{1}{1}{3}
		\emptyPeg{0}{2}{4}
		\filledPeg{1}{2}{5}
		\emptyPeg{2}{2}{6}
		\emptyPeg{0}{3}{7}
		\emptyPeg{1}{3}{8}
		\emptyPeg{2}{3}{9}
		\filledPeg{3}{3}{10}
		\filledPeg{0}{4}{11}
		\emptyPeg{1}{4}{12}
		\filledPeg{2}{4}{13}
		\emptyPeg{3}{4}{14}
		\emptyPeg{4}{4}{15}
	\end{tikzpicture}
	\vDist{}
	
	\begin{tikzpicture}[x=1mm,y=1mm]
		\moveDesc{After $2\to9$ is played:}
		\emptyPeg{0}{0}{1}
		\emptyPeg{0}{1}{2}
		\emptyPeg{1}{1}{3}
		\emptyPeg{0}{2}{4}
		\emptyPeg{1}{2}{5}
		\emptyPeg{2}{2}{6}
		\emptyPeg{0}{3}{7}
		\emptyPeg{1}{3}{8}
		\filledPeg{2}{3}{9}
		\filledPeg{3}{3}{10}
		\filledPeg{0}{4}{11}
		\emptyPeg{1}{4}{12}
		\filledPeg{2}{4}{13}
		\emptyPeg{3}{4}{14}
		\emptyPeg{4}{4}{15}
	\end{tikzpicture}
	\hDist{}
	\begin{tikzpicture}[x=1mm,y=1mm]
		\moveDesc{After $10\to8$ is played:}
		\emptyPeg{0}{0}{1}
		\emptyPeg{0}{1}{2}
		\emptyPeg{1}{1}{3}
		\emptyPeg{0}{2}{4}
		\emptyPeg{1}{2}{5}
		\emptyPeg{2}{2}{6}
		\emptyPeg{0}{3}{7}
		\filledPeg{1}{3}{8}
		\emptyPeg{2}{3}{9}
		\emptyPeg{3}{3}{10}
		\filledPeg{0}{4}{11}
		\emptyPeg{1}{4}{12}
		\filledPeg{2}{4}{13}
		\emptyPeg{3}{4}{14}
		\emptyPeg{4}{4}{15}
	\end{tikzpicture}
	\vDist{}
	
	\begin{tikzpicture}[x=1mm,y=1mm]
		\moveDesc{After $13\to4$ is played:}
		\emptyPeg{0}{0}{1}
		\emptyPeg{0}{1}{2}
		\emptyPeg{1}{1}{3}
		\filledPeg{0}{2}{4}
		\emptyPeg{1}{2}{5}
		\emptyPeg{2}{2}{6}
		\emptyPeg{0}{3}{7}
		\emptyPeg{1}{3}{8}
		\emptyPeg{2}{3}{9}
		\emptyPeg{3}{3}{10}
		\filledPeg{0}{4}{11}
		\emptyPeg{1}{4}{12}
		\emptyPeg{2}{4}{13}
		\emptyPeg{3}{4}{14}
		\emptyPeg{4}{4}{15}
	\end{tikzpicture}
	\vDist{}
	
	This configuration is losing, as there are still multiple pegs on the board, but no more jumps are possible.
	
	\textbf{Alternative, winning play:}
	
	\begin{tikzpicture}[x=1mm,y=1mm]
		\moveDesc{\phantom{After $7\to2$ is played:}}
		\emptyPeg{0}{0}{1}
		\emptyPeg{0}{1}{2}
		\emptyPeg{1}{1}{3}
		\filledPeg{0}{2}{4}
		\filledPeg{1}{2}{5}
		\emptyPeg{2}{2}{6}
		\filledPeg{0}{3}{7}
		\emptyPeg{1}{3}{8}
		\emptyPeg{2}{3}{9}
		\filledPeg{3}{3}{10}
		\filledPeg{0}{4}{11}
		\emptyPeg{1}{4}{12}
		\emptyPeg{2}{4}{13}
		\filledPeg{3}{4}{14}
		\filledPeg{4}{4}{15}
	\end{tikzpicture}
	\hDist{}
	\begin{tikzpicture}[x=1mm,y=1mm]
		\moveDesc{After $7\to2$ is played:}
		\emptyPeg{0}{0}{1}
		\filledPeg{0}{1}{2}
		\emptyPeg{1}{1}{3}
		\emptyPeg{0}{2}{4}
		\filledPeg{1}{2}{5}
		\emptyPeg{2}{2}{6}
		\emptyPeg{0}{3}{7}
		\emptyPeg{1}{3}{8}
		\emptyPeg{2}{3}{9}
		\filledPeg{3}{3}{10}
		\filledPeg{0}{4}{11}
		\emptyPeg{1}{4}{12}
		\emptyPeg{2}{4}{13}
		\filledPeg{3}{4}{14}
		\filledPeg{4}{4}{15}
	\end{tikzpicture}
	\vDist{}

	\begin{tikzpicture}[x=1mm,y=1mm]
		\moveDesc{After $2\to9$ is played:}
		\emptyPeg{0}{0}{1}
		\emptyPeg{0}{1}{2}
		\emptyPeg{1}{1}{3}
		\emptyPeg{0}{2}{4}
		\emptyPeg{1}{2}{5}
		\emptyPeg{2}{2}{6}
		\emptyPeg{0}{3}{7}
		\emptyPeg{1}{3}{8}
		\filledPeg{2}{3}{9}
		\filledPeg{3}{3}{10}
		\filledPeg{0}{4}{11}
		\emptyPeg{1}{4}{12}
		\emptyPeg{2}{4}{13}
		\filledPeg{3}{4}{14}
		\filledPeg{4}{4}{15}
	\end{tikzpicture}
	\hDist{}
	\begin{tikzpicture}[x=1mm,y=1mm]
		\moveDesc{After $15\to6$ is played:}
		\emptyPeg{0}{0}{1}
		\emptyPeg{0}{1}{2}
		\emptyPeg{1}{1}{3}
		\emptyPeg{0}{2}{4}
		\emptyPeg{1}{2}{5}
		\filledPeg{2}{2}{6}
		\emptyPeg{0}{3}{7}
		\emptyPeg{1}{3}{8}
		\filledPeg{2}{3}{9}
		\emptyPeg{3}{3}{10}
		\filledPeg{0}{4}{11}
		\emptyPeg{1}{4}{12}
		\emptyPeg{2}{4}{13}
		\filledPeg{3}{4}{14}
		\emptyPeg{4}{4}{15}
	\end{tikzpicture}
	\vDist{}

	\begin{tikzpicture}[x=1mm,y=1mm]
		\moveDesc{After $6\to13$ is played:}
		\emptyPeg{0}{0}{1}
		\emptyPeg{0}{1}{2}
		\emptyPeg{1}{1}{3}
		\emptyPeg{0}{2}{4}
		\emptyPeg{1}{2}{5}
		\emptyPeg{2}{2}{6}
		\emptyPeg{0}{3}{7}
		\emptyPeg{1}{3}{8}
		\emptyPeg{2}{3}{9}
		\emptyPeg{3}{3}{10}
		\filledPeg{0}{4}{11}
		\emptyPeg{1}{4}{12}
		\filledPeg{2}{4}{13}
		\filledPeg{3}{4}{14}
		\emptyPeg{4}{4}{15}
	\end{tikzpicture}
	\hDist{}
	\begin{tikzpicture}[x=1mm,y=1mm]
		\moveDesc{After $14\!\to\!12$ is played:}
		\emptyPeg{0}{0}{1}
		\emptyPeg{0}{1}{2}
		\emptyPeg{1}{1}{3}
		\emptyPeg{0}{2}{4}
		\emptyPeg{1}{2}{5}
		\emptyPeg{2}{2}{6}
		\emptyPeg{0}{3}{7}
		\emptyPeg{1}{3}{8}
		\emptyPeg{2}{3}{9}
		\emptyPeg{3}{3}{10}
		\filledPeg{0}{4}{11}
		\filledPeg{1}{4}{12}
		\emptyPeg{2}{4}{13}
		\emptyPeg{3}{4}{14}
		\emptyPeg{4}{4}{15}
	\end{tikzpicture}
	\vDist{}

	\begin{tikzpicture}[x=1mm,y=1mm]
		\moveDesc{After $11\to13$ is played:}
		\emptyPeg{0}{0}{1}
		\emptyPeg{0}{1}{2}
		\emptyPeg{1}{1}{3}
		\emptyPeg{0}{2}{4}
		\emptyPeg{1}{2}{5}
		\emptyPeg{2}{2}{6}
		\emptyPeg{0}{3}{7}
		\emptyPeg{1}{3}{8}
		\emptyPeg{2}{3}{9}
		\emptyPeg{3}{3}{10}
		\emptyPeg{0}{4}{11}
		\emptyPeg{1}{4}{12}
		\filledPeg{2}{4}{13}
		\emptyPeg{3}{4}{14}
		\emptyPeg{4}{4}{15}
	\end{tikzpicture}
	\vDist{}

	This configuration only has a single peg remaining and is therefore winning.
	\end{toappendix}

\subsubsection{Example: Misrouted Train}

We modelled Dresden Central Station to analyse a misrouted train. The train was supposed to arrive at Platform~$12$ but instead arrived at Platform~$13$. Our goal is to determine how much responsibility each switch in the train station has for this misrouting. The relevant fragment of the station is depicted in Figure~\ref{fig:dresden_hbf}. The path taken by the train is indicated in bold.

\newcommand{\trackLdotsHOffset}{0.6cm}
\newcommand{\trackLdotsVOffset}{-0.15cm}
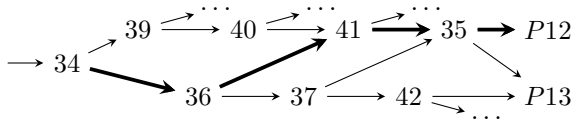
\begin{figure}[H]
	\centering
	\begin{tikzpicture}[>=stealth,node distance=1.4cm,auto, 
		state/.style={circle,inner sep=2pt}]
		\node[state,initial,initial text=]  (34) {$34$};
		\node[state] (39) [above right = 0.0cm and 0.5cm of 34] {$39$};
		\node[state] (40) [right of = 39] {$40$};
		\node[state] (41) [right of = 40] {$41$};
		\node[state] (36) [below right= 0.0cm and 1.3cm of 34] {$36$};
		\node[state] (37) [right of = 36] {$37$};
		\node[state] (35) [right of = 41] {$35$};
		\node[state] (42) [right of = 37] {$42$};
		\node[state] (55) [right = 0.5cm of 35] {$P12$};
		\node[state] (56) [right = 1.1cm of 42] {$P13$};
		\node[state] (dots1) [above right=\trackLdotsVOffset and \trackLdotsHOffset of 39] {$\ldots$};
		\node[state] (dots2) [above right=\trackLdotsVOffset and \trackLdotsHOffset of 40] {$\ldots$};
		\node[state] (dots3) [above right=\trackLdotsVOffset and \trackLdotsHOffset of 41] {$\ldots$};
		\node[state] (dots4) [below right=\trackLdotsVOffset and \trackLdotsHOffset of 42] {$\ldots$};
		\path[->]
		(34) edge node {} (39)
		(39) edge node {} (40)
		(40) edge node {} (41)
		(41) edge node {} (35)
		(35) edge node {} (55)
		(35) edge node {} (56)
		(34) edge node {} (36)
		(36) edge node {} (41)
		(36) edge node {} (37)
		(37) edge node {} (35)
		(37) edge node {} (42)
		(42) edge node {} (56)
		(39) edge node {} (dots1)
		(40) edge node {} (dots2)
		(41) edge node {} (dots3)
		(42) edge node {} (dots4);
		\path[->, line width=0.5mm]
		(34) edge node {} (36)
		(36) edge node {} (41)
		(41) edge node {} (35)
		(35) edge node {} (55);
	\end{tikzpicture}
	\caption{A train misrouted to Platform $P12$ instead of $P13$ in Dresden Central Station and the path the train took.}
	\label{fig:dresden_hbf}
\end{figure}

As we are interested in a quantitative analysis of responsibility, we choose pessimistic responsibility. This yields four responsible states: $35$ has responsibility $3/4$, whereas $36$, $37$ and $42$ have responsibility $1/12$ each. The reason for $35$'s high responsibility is that it can route the train to the correct track without any cooperation from other switches. On the other hand, $36$, $37$ and $42$ need to all be in the correct position to achieve the same result. As the default assumption in pessimistic responsibility is that other switches do not cooperate, this three-way cooperation only occurs in few coalitions and therefore, the total responsibility of the three states is lower than that of $35$.

\newcommand{\trainPath}{\!\to\!}
Another possible route to Platform~$13$ is $34\trainPath39\trainPath40\trainPath41\trainPath35\trainPath P13$. Note however that $34, 39, 40\text{ and }41$ have responsibility $0$. This is because the route relies on $35$ cooperating. As $35$ can route the train correctly on its own, there is no benefit in also changing the route at the other switches.

\subsection{Stochastic Algorithm}

\begin{table*}[ht]
	\centering
	{ \small
		\begin{tabular}{l r | r r r r r r}
			& & \multicolumn{2}{c}{$t=1s$} & \multicolumn{2}{c}{$t=10s$} & \multicolumn{2}{c}{$t=60s$} \\
			name & states & \multicolumn{1}{c}{$n$} & \multicolumn{1}{c}{$\sigma$} & \multicolumn{1}{c}{$n$} & \multicolumn{1}{c}{$\sigma$} & \multicolumn{1}{c}{$n$} & \multicolumn{1}{c}{$\sigma$} \\
			\hline
			\texttt{alternating\_bit} & $297$ & $38.4$k & $0.0304$ & $335.1$k & $0.0091$ & $1844.0$k & $0.0038$ \\
			\texttt{brp}, $N = 4$, $\mathit{MAX} = 2$ & $173$ & $685.4$k & $0.0015$ & $6963.9$k & $0.0005$ & $41754.6$k & $0.0002$ \\
			\texttt{brp}, $N = 16$, $\mathit{MAX} = 3$ & $886$ & $71.2$k & $0.0087$ & $663.5$k & $0.0028$ & $3826.8$k & $0.0011$ \\
			\texttt{crowds}, $\mathit{TR} = 3$, $\mathit{CS} = 5$ & $1198$ & $49.3$k & $0.0120$ & $372.3$k & $0.0041$ & $2177.8$k & $0.0018$ \\
			\texttt{dining\_philosophers}, $N=3$ & $36$ & $1606.6$k & $0.0015$ & $16145.9$k & $0.0005$ & $95952.8$k & $0.0002$ \\
			\texttt{dining\_philosophers}, $N=5$ & $393$ & $35.5$k & $0.0135$ & $321.9$k & $0.0049$ & $1865.5$k & $0.0022$ \\
			\texttt{dresden\_railways} & $54$ & $1286.8$k & $0.0009$ & $12537.2$k & $0.0003$ & $76475.9$k & $0.0001$ \\
			\texttt{generals}, $N=3$ & $20$ & $2055.2$k & $0.0017$ & $20655.9$k & $0.0005$ & $117232.9$k & $0.0002$ \\
			\texttt{generals}, $N=5$ & $112$ & $50.8$k & $0.0166$ & $603.8$k & $0.0049$ & $3626.3$k & $0.0019$ \\
			\texttt{generals}, $N=8$ & $1280$ & ($7.9$k) & ($0.0909$) & $69.9$k & $0.0242$ & $313.6$k & $0.0104$ \\
		\end{tabular}
	}
	
	\caption{Evaluation of the stochastic algorithm, where $n$ is the number of samples per run and $\sigma$ is the standard deviation of the result from each run from the reference value.}
	\label{tab:sampling_results}
\end{table*}

The stochastic algorithm uses the following transformation.
\begin{align*}
    \semivalue(\coopGame, s) = & \!\!\!\! \sum_{C \subseteq S \setminus \{s\}} \weights_{\left|C\right|}(\semivalue) (\coopGame(C \cup \{s \}) - \coopGame(C)) \\
    = & \sum_{i=0}^{n-1} \weights_i(\semivalue) \cdot \!\!\!\!\! \sum_{\substack{C \subseteq S \setminus \{s \},\\|C| = i}} \!\!\!\! (\coopGame(C \cup \{s\}) - \coopGame(C)).
\end{align*}
As each summand of the inner sum is either $0$ or $1$, it is sufficient to count for how many coalitions the summand is $1$. The stochastic algorithm estimates this count by uniformly sampling coalitions of size $i$. If $x$ out of $y$ coalitions have value $1$, then the expected count for size $i$ is $(x/y) \cdot \binom{n}{i}$. As long as each size is sampled at least once, the algorithm is an unbiased estimator for $\semivalue(\coopGame, s)$.

The number of samples for size $i$ should be proportional to $\binom{n}{i} \cdot \weights_i(\semivalue)$. In the case of the Shapley value, each size should therefore get the same number of samples.

We have evaluated this on a collection of benchmarks. \texttt{alternating\_bit} is a simple message transition protocol, \texttt{brp} is a protocol to transfer files consisting of $N$ chunks (with $\textit{MAX}$ attempts), \texttt{crowds} models anonymised message routing through a group of size $\mathit{CS}$ (with $\textit{TR}$ total runs), \texttt{dining\_philosophers} models the well-known dining philosophers problem, \texttt{dresden\_railways} models the switches in Dresden Central Station, where a train has to be routed to a specific platform and \texttt{generals} models the $n$-generals problem, who have to decide independently whether to attack or not.%
The benchmarks \texttt{brp} and \texttt{crowds} are taken from \citet{PRISMBenchmarkSuite}.

For each model, we have sampled for $t$ seconds (where $t \in \{1, 10, 60\}$) and then computed the responsibility of all states using these samples. To evaluate the quality of our samples, we have repeated this procedure $20$ times and determined the standard deviation of the estimated responsibility from the actual value\footnote{For the larger models, we cannot compute the exact responsibility. We therefore used the average of all runs with $t = 60$ as reference value, as this is the best approximation we have.}. Table~\ref{tab:sampling_results} presents the results. The number of states of each model is given and for each run, $n$ indicates the average number of samples and $\sigma$ the standard deviation of the samples from the reference value.

Parentheses indicate insufficient coverage, which we detect by analysing the sum of the responsibilities. If it is consistently much smaller than $1$, coverage is likely insufficient.

As expected, the standard deviation decreases if we sample for longer. Furthermore, bigger models show a bigger deviation than smaller models for the same sampling duration. This is to be expected as each sample takes longer for bigger models.

\subsection{State Grouping}
State grouping works by partitioning the set of states $S$ into groups $G_1, \ldots, G_m$. Instead of analysing all subsets $C \subseteq S$, we instead analyse $C = \bigcup_{i \in I} G_i$ for all $I \subseteq \{1, \ldots, m \}$, i.e. if two states are in the same state group, they are either both included in the coalition or neither of them is.

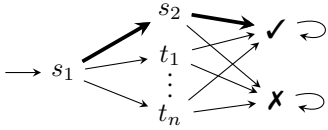
\begin{figure}[H]
	\centering
	\begin{tikzpicture}[>=stealth,node distance=1.4cm,auto,state/.style={circle,inner sep=2pt}]
		\node[state,initial,initial text=]  (one) {$s_1$};
		\node[state] (two) [above right = 0.4cm and 1cm of one] {$s_2$};
		\node[state] (t1) [below right = -0.65cm and 1cm of one] {$t_1$};
		\node (tdots) [below = -0.4cm of t1] {\vdots};
		\node[state] (tn) [below = -0.2cm of  tdots] {$t_n$};
		\node[state] (destination) [below right = -0.18cm and 1cm of two] {\cmark};
		\node[state] (accident) [below = 0.4cm of destination] {\xmark};
		
		\path[->]
		(one) edge node {} (two)
		(one) edge node {} (t1)
		(one) edge node {} (tn)
		(two) edge node {} (destination)
		(two) edge node {} (accident)
		(t1) edge node {} (destination)
		(t1) edge node {} (accident)
		(tn) edge node {} (destination)
		(tn) edge node {} (accident)
		(destination) edge [loop right] node {} (destination)
		(accident) edge [loop right] node {} (accident);
		\path[->, line width=0.5mm]
		(one) edge node {} (two)
		(two) edge node {} (destination);
		
	\end{tikzpicture}
	\caption{Railway network similar to that from Figure~\ref{fig:train_example}, but with multiple switches $t_1, \ldots, t_n$.}
	\label{fig:train_example_groups}
\end{figure}

To give an intuition for state groups, recall the example from Figure~\ref{fig:train_example}. We can increase the complexity of the model by replacing $s_3$ with several switches $t_1, \ldots, t_n$, as shown in Figure~\ref{fig:train_example_groups}. Running our tool for $n=5$ reveals that each $t_i$ has responsibility $0.0238$, whereas $s_1$ has a responsibility of $0.3571$ and $s_2$ has a responsibility of $0.5238$ (all values rounded to four decimal places). If we do not want to analyse individual responsibility of the switches $t_1$ to $t_5$ (perhaps because they are switched by the same controller, or because we want to reduce the complexity of the model), we can use the state groups $\{s_1\}, \{s_2\}, \{t_1, \ldots, t_5\}$. With these state groups, the responsibilities are now equal to those in Table~\ref{tab:example_resps}, i.e. $\{s_1\}$ has responsibility $1/6$, $\{s_2\}$ has responsibility $2/3$ and $\{t_1, \ldots, t_5\}$ has responsibility $1/6$.

In addition to simplifying the results, state groups also decrease runtime, as the algorithm is exponential only in the number of \emph{groups}, but linear in the number of states.

\subsubsection{Example: Dining Philosophers} Consider an implementation of the dining philosophers with a simple scheduler. For four philosophers, this model has $180$ states, which cannot be analysed with the exact algorithm. If we group the states by whose turn it is, we can compute responsibility values in less than a millisecond (and it is revealed that, in our case, all philosophers are equally responsible for the deadlock).

\subsubsection{Example: BRP} It is also possible to combine the stochastic algorithm with state grouping. For example, consider the \texttt{brp} model from Table~\ref{tab:sampling_results}. If we increase the parameters to $N=64, \mathit{MAX}=5$, the model has $5192$ states and even $60$s of sampling result in insufficient coverage for individual responsibility (as the sum of responsibilities is consistently below $1$). If we instead group states by which chunk is being processed, we get 64 groups and after running for $1$s, the tool is able to approximate the responsibility values with a standard deviation of $0.0006$.

\section{Conclusion}
\label{sec:conclusion}

We have presented two notions of backward responsibility. For optimistic responsibility, every state with positive responsibility has the same amount of responsibility. The characterisation provides a simple test to find the states with positive responsibility by computing whether they can change the outcome by themselves. While this makes computation straight-forward, it also means that it does not rank states by responsibility in the way pessimistic responsibility does.

Pessimistic responsibility, on the other hand, is harder to compute, but can give positive responsibility both to states on the counterexample and to states that are not. Furthermore, not all responsible states have the same responsibility.

We have demonstrated that our technique works in practice and that a stochastic algorithm can be used to analyse much larger models than would otherwise be feasible.

\textbf{Future Work:} Our investigation was restricted to safety games. As our definitions can readily be adapted to other classes of games, such as Büchi  games, it is of interest to determine whether our results still hold for these classes and to analyse the complexity of the algorithms for them.

Another avenue for future work is the presentation of the data to the user. Our tool gives the responsibility values for each state. To facilitate debugging, it would be useful to take this state-based responsibility and map it back to the specification language (in our case, this is the \Prism{} language). However, such a mapping is non-trivial, as there is no direct correspondence between states and source-code lines.

\section*{Acknowledgements}

Funded by the German Research Foundation (DFG, Deutsche Forschungsgemeinschaft) as part of Germany’s Excellence Strategy -- EXC 2050/1 -- Project ID 390696704 -- Cluster of Excellence ``Centre for Tactile Internet with Human-in-the-Loop'' (CeTI) of Technische Universität Dresden, by DFG Grant 389792660 as part of TRR 248 (Foundations of Perspicuous Software Systems) and by BMBF (Federal Ministry of Education and Research) in DAAD project 57616814 (SECAI, School of Embedded Composite AI) as part of the program Konrad Zuse Schools of Excellence in Artificial Intelligence.
\bibliography{aaai24}

\end{document}